\documentclass[
     final            
  ]
  {aipproc}

\layoutstyle{8x11double}
\usepackage{amsmath}

\begin{document}

\title{Optical Conductivity of Metals from First Principles}

\classification{71.15.Qe,71.45.Gm,72.15.Eb}
\keywords      {Optical conductivity, Dielectric function, Random-phase approximation, Interband transitions}

\author{Arno Schindlmayr}{
  address={Department Physik, Universit\"at Paderborn, 33095 Paderborn, Germany}
}

\begin{abstract}
A computational method to obtain optical conductivities from first principles is presented. It exploits a relation between the conductivity and the complex dielectric function, which is constructed from the full electronic band structure within the random-phase approximation. In contrast to the Drude model, no empirical parameters are used. As interband transitions as well as local-field effects are properly included, the calculated spectra are valid over a wide frequency range. As an illustration I present quantitative results for selected simple metals, noble metals, and ferromagnetic transition metals. The implementation is based on the full-potential linearized augmented-plane-wave method.
\end{abstract}

\maketitle
\setcounter{page}{157}


\section{Introduction}
\thispagestyle{plain}

The optical conductivity $\sigma$ yields a wealth of information about the electrical conductance properties of a material and its reaction to applied electromagnetic fields in linear response. It is formally defined by the ratio
\begin{equation}\label{Eq:current}
\mathbf{j}(\mathbf{q},\omega) = \sigma(\mathbf{q},\omega) \mathbf{E}(\mathbf{q},\omega)
\end{equation}
between the total electrical field $\mathbf{E}$ inside the material and the induced current density $\mathbf{j}$. As all quantities are taken as macroscopic averages over many unit cells, Eq.\ (\ref{Eq:current}) is only valid for small wave vectors $\mathbf{q}$, otherwise the variations at the atomic scale must be explicitly included. In general, the conductivity is a tensor, but it reduces to a scalar for systems with cubic symmetry.

The conductivity can be properly expressed within a quantum-mechanical context through the Kubo formula \cite{Kubo1957}, but most actual studies of the optical and transport properties of solids are based on simpler schemes, such as the classical Drude model \cite{Drude1900}. While its simplicity makes the Drude model very attractive, there are also clear disadvantages. In particular, it relies on parameters, especially the phenomenological relaxation time, that are difficult to establish unambiguously, because the optimal values may depend on the frequency range and on the specific optical function under consideration. Besides, the original Drude model ignores interband transitions in real inhomogeneous solids, which already dominate in the visible range, so that extensions with even more empirical parameters become necessary in this case.

Here I describe a parameter-free \textit{ab initio} method for calculating the optical conductivity of metals over a wide frequency range, which takes interband transitions and local-field effects in inhomogeneous solids into account. Quantitative results are shown for selected simple metals, noble metals and transition metals from the infrared to the ultraviolet region of the spectrum.

\section{Theoretical framework}
\thispagestyle{plain}

Due to the small momentum transfer in optical absorption it is usually sufficient to consider the limit $\mathbf{q} \to \mathbf{0}$. The basic equation used in this work is the relation
\begin{equation}\label{Eq:basic}
\varepsilon_\mathrm{M}(\omega) = 1 + \frac{4 \pi i}{\omega} \lim_{\mathbf{q} \to \mathbf{0}} \sigma(\mathbf{q},\omega)
\end{equation}
between the conductivity in the long-wave-length limit and the macroscopic dielectric function, which follows from Maxwell's equations. The latter is defined as \cite{Adler1962,Wiser1963}
\begin{equation}\label{Eq:macroscopic}
\varepsilon_\mathrm{M}(\omega) = \lim_{\mathbf{q} \to \mathbf{0}} \frac{1}{\varepsilon^{-1}(\mathbf{q},\omega)}
\end{equation}
in terms of the matrix inverse of the microscopic dielectric function projected onto a plane-wave basis
\begin{equation}
\varepsilon^{-1}(\mathbf{q},\omega) = \int \varepsilon^{-1}(\mathbf{r},\mathbf{r}';\omega) e^{-i \mathbf{q} \cdot (\mathbf{r} - \mathbf{r}')} \,d^3r \,d^3r' \;.
\end{equation}
However, actual applications often replace the left-hand side of (\ref{Eq:basic}) by $\lim_{\mathbf{q} \to \mathbf{0}} \varepsilon(\mathbf{q},\omega)$, which is computationally simpler, because it requires only a single element instead of the entire matrix, but neglects local-field effects.

The microscopic dielectric function represents the link to the electronic structure of the material and can be rigorously derived in time-dependent density-functional theory \cite{Petersilka1996,Botti2007}. Here I use the random-phase approximation
\begin{equation}\label{Eq:RPA}
\varepsilon(\mathbf{r},\mathbf{r}';\omega) = \delta(\mathbf{r} - \mathbf{r}') - \int v(\mathbf{r}-\mathbf{r}'') \chi_0(\mathbf{r}'',\mathbf{r}';\omega) \,d^3r''
\end{equation}
with the Coulomb potential $v(\mathbf{r} - \mathbf{r}') = 1 / | \mathbf{r} - \mathbf{r}' |$ and the susceptibility (density correlation function) of the non-interacting auxiliary Kohn-Sham electrons
\begin{align}
\chi_0(\mathbf{r},\mathbf{r};\omega) = {} & \lim_{\eta \to 0} \sum_\sigma \sum_{n,\mathbf{k}} \sum_{n',\mathbf{k}'} \left( f_{n'\mathbf{k}'\sigma} - f_{n\mathbf{k}\sigma} \right) \label{Eq:susceptibility}\\
& \times \frac{\varphi_{n\mathbf{k}\sigma}(\mathbf{r}) \varphi^*_{n'\mathbf{k}'\sigma}(\mathbf{r}) \varphi_{n'\mathbf{k}'\sigma}(\mathbf{r}') \varphi^*_{n\mathbf{k}\sigma}(\mathbf{r}')}{\omega - \epsilon_{n\mathbf{k}\sigma} + \epsilon_{n'\mathbf{k}'\sigma} + i \eta} \;.\nonumber
\end{align}
The sums run over all single-particle states with band number $n$, wave vector $\mathbf{k}$, and spin $\sigma$, which are obtained from a self-consistent solution of the Kohn-Sham equation \cite{Kohn1965} of stationary density-functional theory \cite{Hohenberg1964}. The occupation numbers $f_{n\mathbf{k}\sigma}$ obey the Fermi distribution.

While the method itself can be formulated in a very general way, it should be pointed out that the present implementation has some restrictions: The random-phase approximation (\ref{Eq:RPA}) only includes the scattering between the electrons and the influence of the static crystal potential on the electronic structure. This is appropriate for most optical experiments, but phonon scattering cannot be neglected in studies of the electrical resistivity. In such cases the effects of dynamical lattice vibrations on the dielectric function must be explicitly taken into account \cite{Allen1988}. Furthermore, the construction (\ref{Eq:susceptibility}) assumes a perfect periodic crystal, where the Bloch wave vector $\mathbf{k}$ is a good quantum number, without defects or disorder.

\section{Results}
\thispagestyle{plain}

\begin{figure}[b!]
\includegraphics*[width=\columnwidth]{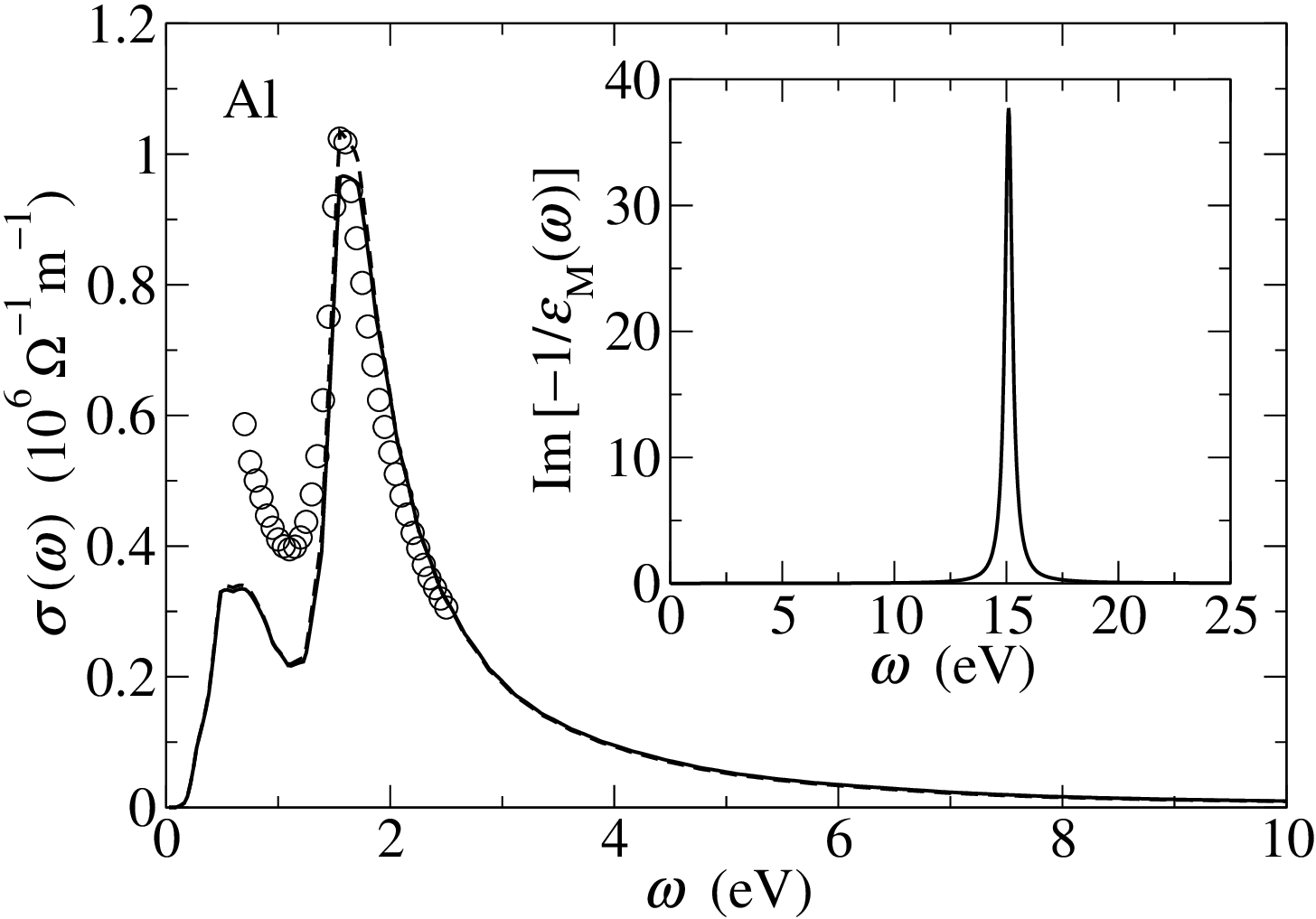}
\caption{Optical conductivity of aluminum compared to experimental data from Ref.\ \cite{Mathewson1971} (circles). In all figures the solid and dashed line indicate results with and without local-field effects, respectively. The inset shows the energy-loss function, which is dominated by the plasmon resonance near 15\,eV.}
\label{Fig:sigma-al}
\end{figure}

For the calculations I employ the full-potential linearized augmented-plane-wave method. Exchange and correlation are treated in the local-density approximation \cite{Kohn1965}. The Kohn-Sham orbitals are expanded in plane waves with reciprocal lattice vectors up to $G_\mathrm{max} = 4\,\mathrm{Bohr}^{-1}$ in the interstitial region and up to angular momentum $l_\mathrm{max} = 10$ inside the muffin-tin spheres. I include second energy derivatives of the muffin-tin functions in order to improve the description of higher unoccupied states \cite{Friedrich2006}. The mixed product basis for the response functions and the Coulomb matrix is constructed with $G'_\mathrm{max} = 5\,\mathrm{Bohr}^{-1}$ and $L_\mathrm{max} = 4$. Integrations over the Brillouin zone are performed with 32$\times$32$\times$32 mesh points, the singularity of the Coulomb matrix at the zone center $\mathbf{q} = \mathbf{0}$ is treated analytically \cite{Friedrich2009}. All spectra are calculated at the experimental lattice constants and at zero temperature.

In the following I present results for selected metallic systems. As a first example, Fig.\ \ref{Fig:sigma-al} displays the real part of the optical conductivity of a simple metal, aluminum, whose electronic structure resembles that of the homogeneous electron gas. Nevertheless, interband transitions play an important role in the visible range and give rise to the resonances at 0.5\,eV and 1.5\,eV. Both are predicted by this theoretical approach in very good agreement with experimental data from ellipsometry measurements \cite{Mathewson1971}. Local-field effects have only a very small influence. The inset displays another independent optical quantity, the so-called energy-loss function $\mathop{\mathrm{Im}} [ -1 / \varepsilon_\mathrm{M}(\omega) ]$, which is dominated by the plasmon resonance near 15\,eV.

\begin{figure}[t!]
\includegraphics*[width=\columnwidth]{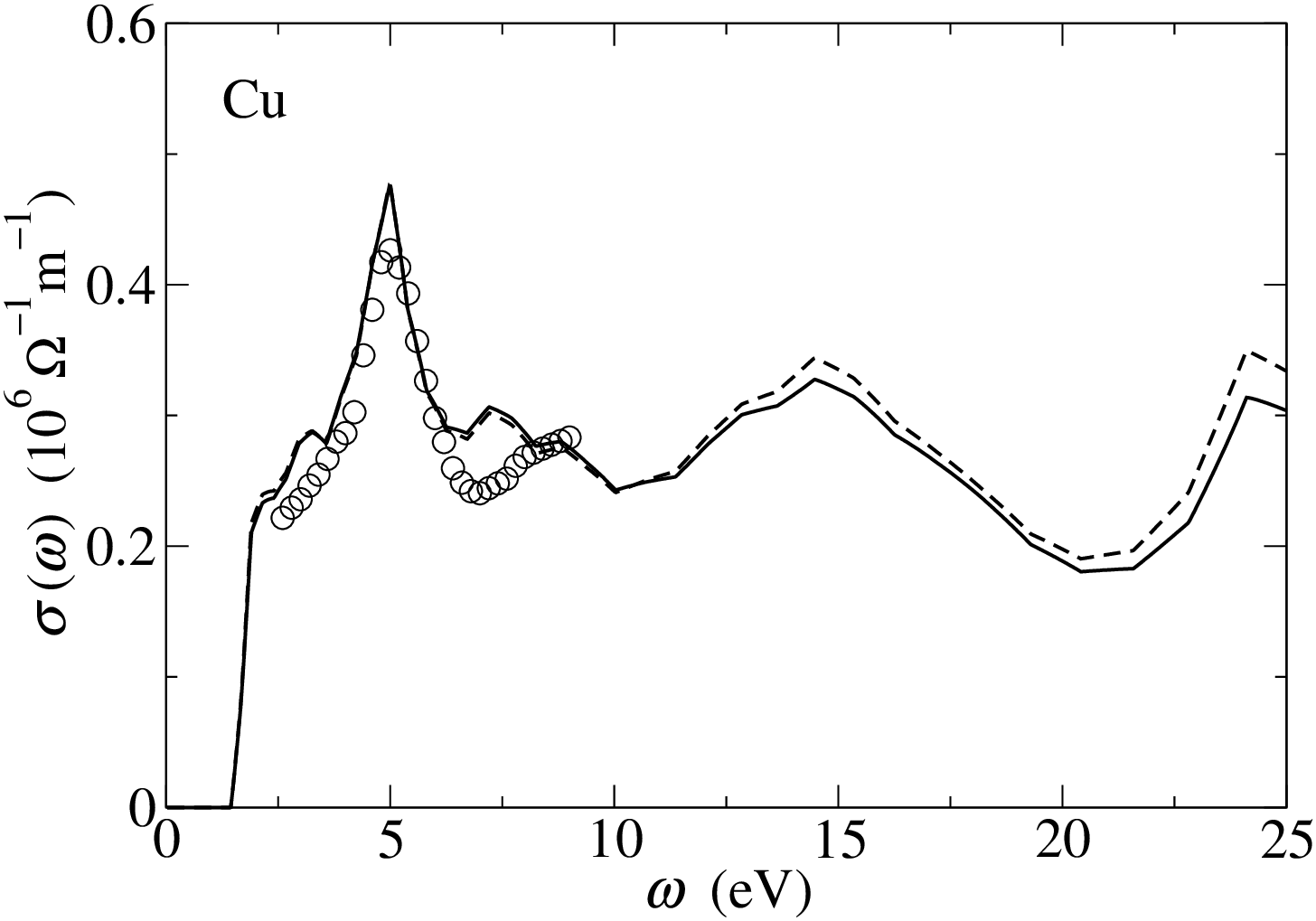}
\caption{Optical conductivity of copper compared to experimental data from Ref.\ \cite{Stahrenberg2001} (circles).}
\label{Fig:sigma-cu}
\end{figure}

\begin{figure}[t!]
\includegraphics*[width=\columnwidth]{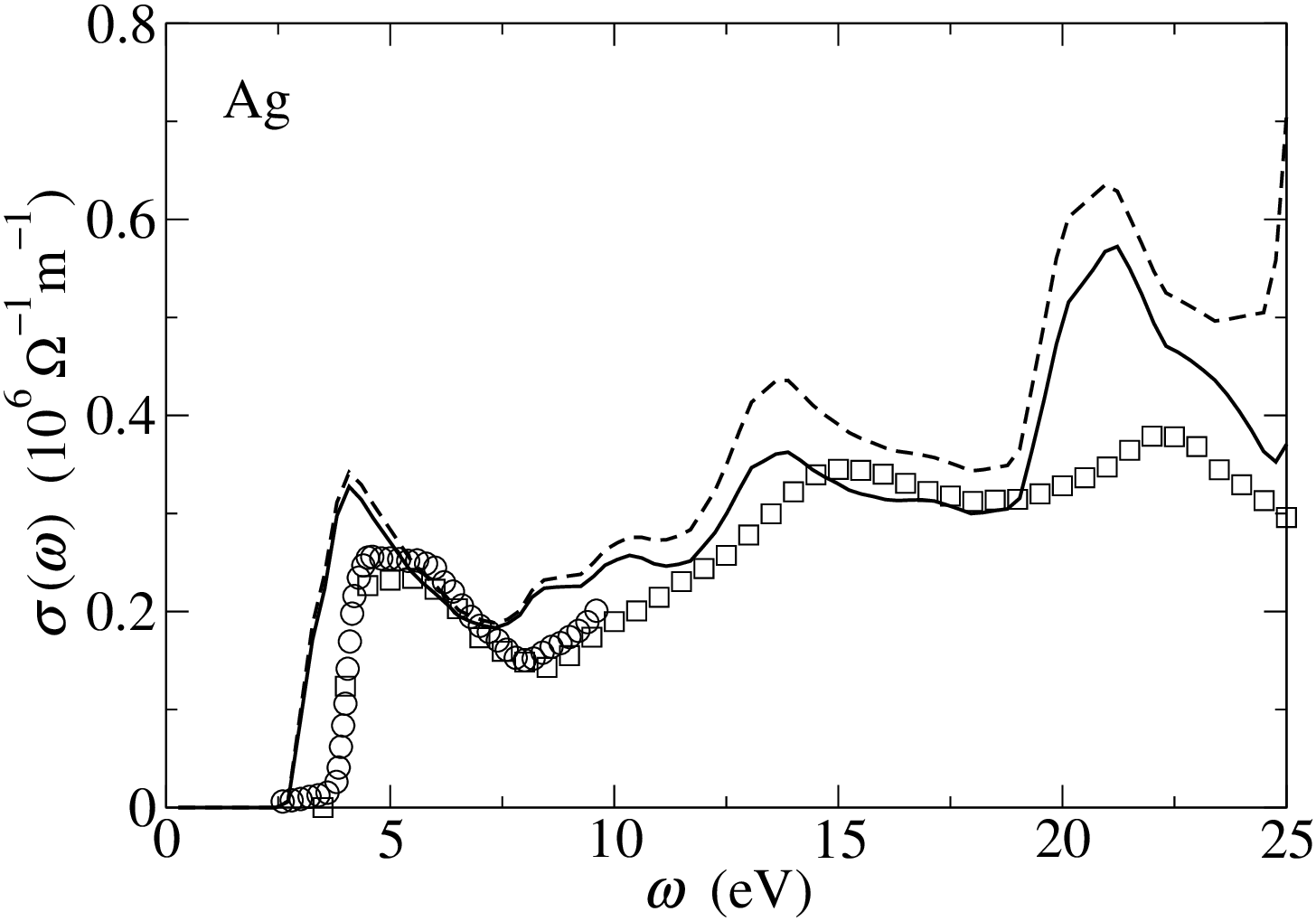}
\caption{Optical conductivity of silver compared to experimental data from Refs.\ \cite{Stahrenberg2001} (circles) and \cite{Leveque1983} (squares).}
\label{Fig:sigma-ag}
\end{figure}

In comparison, the spectral features of noble metals exhibit more structure, as shown in Fig.\ \ref{Fig:sigma-cu} for copper and in Fig.\ \ref{Fig:sigma-ag} for silver. In these cases transitions from the occupied $d$ bands dominate in the considered energy window and give rise to pronounced peaks in the visible to ultraviolet region when threshold values for transitions are reached. All the features measured by experimental spectroscopies \cite{Stahrenberg2001,Leveque1983} are again very well reproduced.

\begin{figure}[t!]
\includegraphics*[width=\columnwidth]{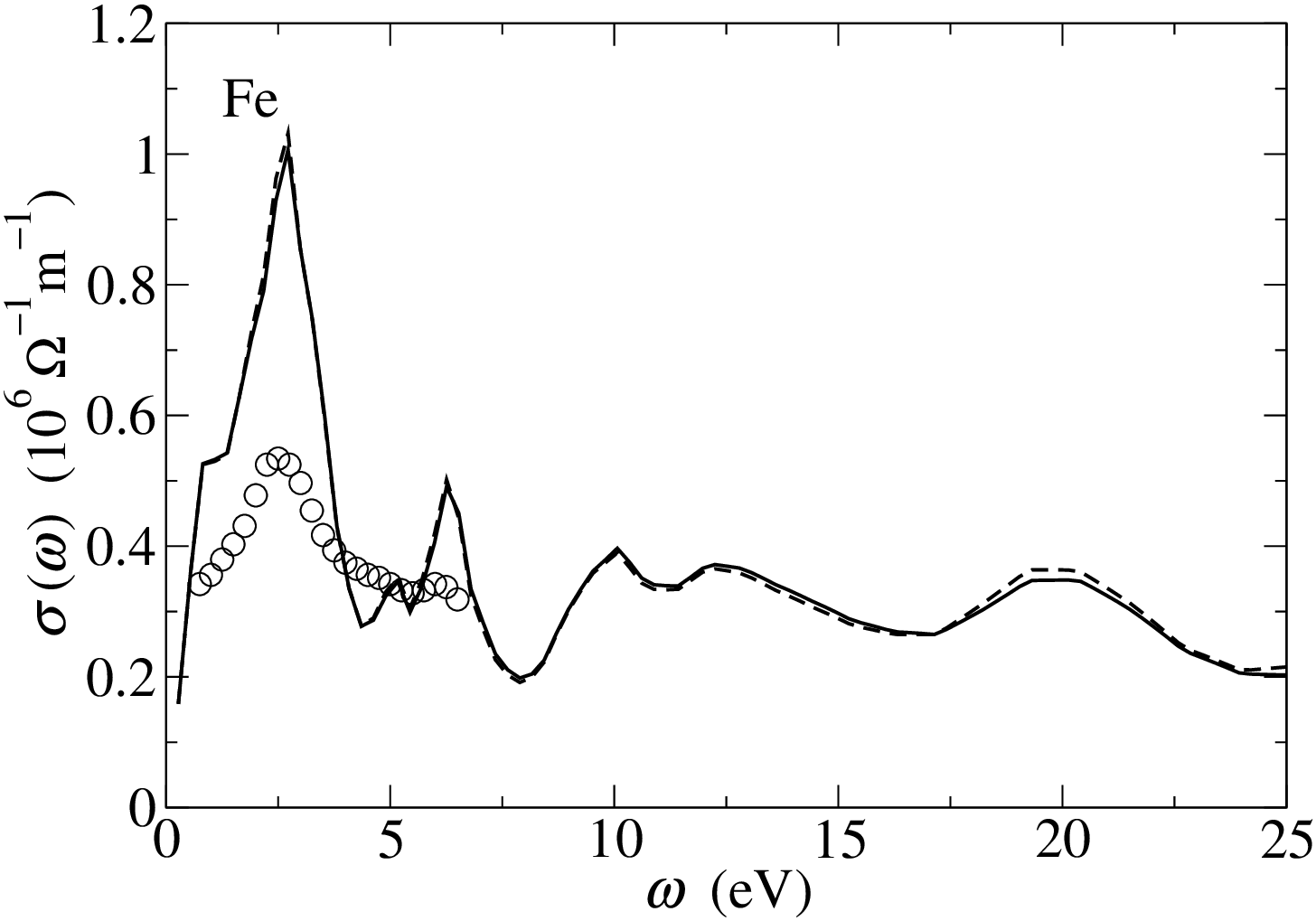}
\caption{Optical conductivity of iron compared to experimental data from Ref.\ \cite{Johnson1974} (circles).}
\label{Fig:sigma-fe}
\end{figure}

\begin{figure}[t!]
\includegraphics*[width=\columnwidth]{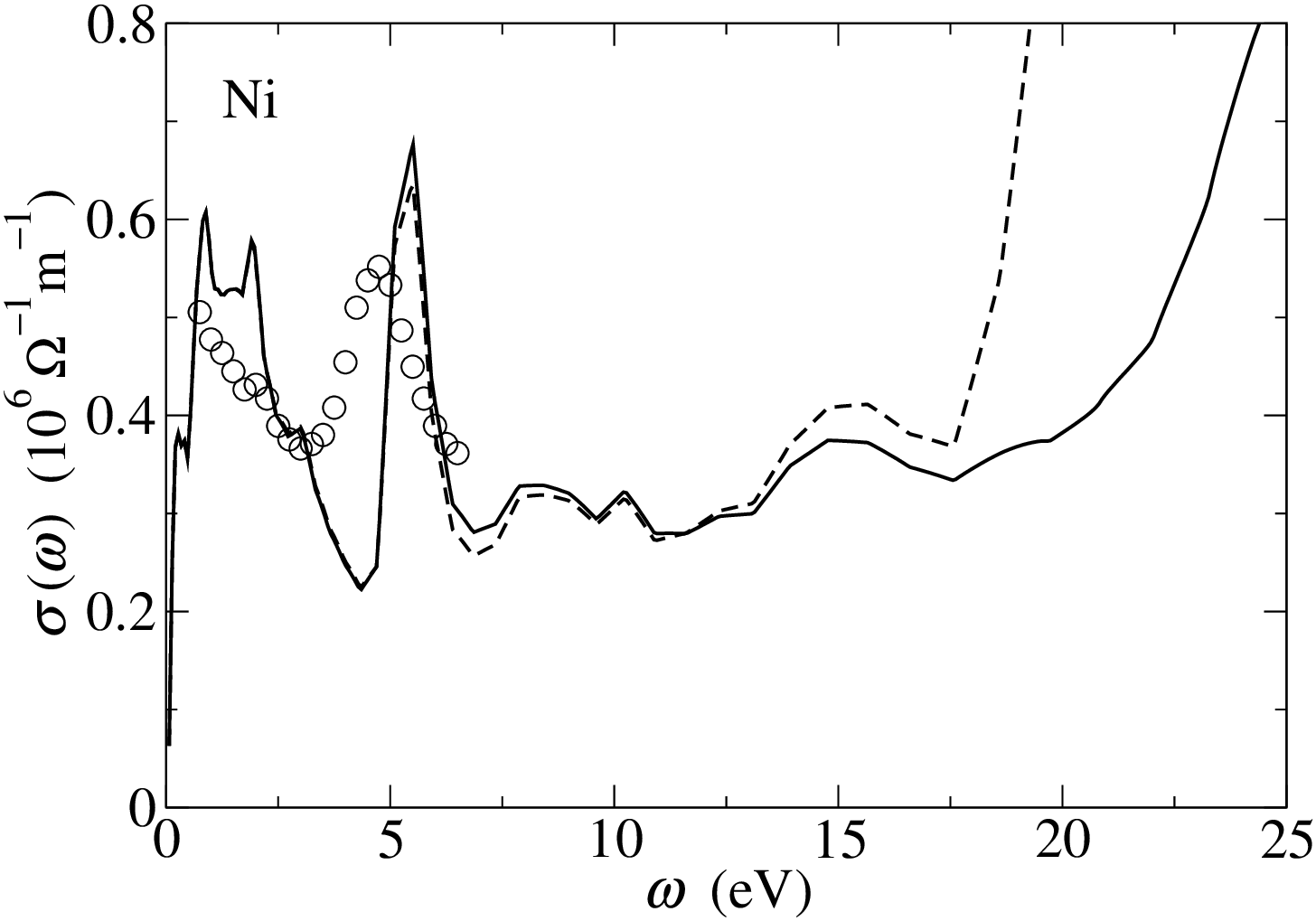}
\caption{Optical conductivity of nickel compared to experimental data from Ref.\ \cite{Johnson1974} (circles).}
\label{Fig:sigma-ni}
\end{figure}

As shown in Fig.\ \ref{Fig:sigma-fe} for iron and in Fig.\ \ref{Fig:sigma-ni} for nickel, the spectra of transition metals are even more complex due to their complicated electronic structure. In these materials interband transitions between the partially occupied and weakly dispersive $d$ bands near the Fermi level give rise to a prominent peak structure at lower frequencies than in the noble metals. Furthermore, the exchange splitting in ferromagnets effectively doubles the band structure, leading to distinct optical transitions in the majority and minority spin channels that create additional features in the spectra. The theoretical curves are in good agreement with optical measurements \cite{Johnson1974}, but they have sharper features, as no artificial broadening is applied.

\section{Conclusions}
\thispagestyle{plain}

I have described a computational method for determining optical conductivities without any empirical parameters from first principles. Interband transitions and local-field effects are fully taken into account. Quantitative results for selected simple metals, noble metals, and transition metals confirm that practical calculations are feasible and that the method yields results in good quantitative agreement with experiments for materials with very different electronic characteristics. The essential spectral features are accurately described over a wide frequency range. Other optical quantities, such as the refractive index, the energy-loss function, or the reflectivity, can be derived in the same fashion from the complex dielectric function. Although the large numerical effort precludes a straightforward application to microscale device structures, such atomistic simulations of materials properties can be used to determine parameters for simplified models like the Drude model and its various extensions.


\begin{theacknowledgments}
Financial support from the Deutsche Forschungsgemeinschaft through the Priority Programme 1145 is gratefully acknowledged.
\end{theacknowledgments}



\bibliographystyle{aipproc}   


%


\end{document}